# Antiferromagnetic phase of the gapless semiconductor V$_3$Al

M.E. Jamer[1], B.A. Assaf[1], G.E. Sterbinsky[2], D. Arena[2], L.H. Lewis[3], A.A. Saúl[4, 5], G. Radtke[6] and D. Heiman[1]

[1]Department of Physics, Northeastern University, Boston, MA 02115 USA
[2] Photon Sciences Directorate, Brookhaven National Laboratory, Upton, NY, 11973 USA
[3]Department of Chemical Engineering, Northeastern University, Boston, MA 02115 USA
[4]Aix-Marseille Université, CINaM-CNRS UMR 7325 Campus de Luminy, 13288 Marseille Cedex 9, France
[5]Department of Civil and Environmental Engineering, MIT, Cambridge MA 02139 USA
[6]Institut de Minéralogie, de Physique des Matériaux et de Cosmochimie (IMPMC) Sorbonne Universités - UPMC Univ Paris 06, UMR CNRS 7590, Muséum National d'Histoire Naturelle, IRD UMR 206, 4 Place Jussieu, F-75005 Paris, France

Discovering new antiferromagnetic compounds is at the forefront of developing future spintronic devices without fringing magnetic fields. The antiferromagnetic gapless semiconducting D0$_3$ phase of V$_3$Al was successfully synthesized via arc-melting and annealing. The antiferromagnetic properties were established through synchrotron measurements of the atom-specific magnetic moments, where the magnetic dichroism reveals large and oppositely-oriented moments on individual V atoms. Density functional theory calculations confirmed the stability of a type G antiferromagnetism involving only two-third of the V atoms, while the remaining V atoms are nonmagnetic. Magnetization, x-ray diffraction and transport measurements also support the antiferromagnetism. This archetypal gapless semiconductor may be considered as a cornerstone for future spintronic devices containing antiferromagnetic elements.

## I. INTRODUCTION

Spin gapless semiconductors (SGS) are novel materials that combine both spin-polarized carriers and novel semiconducting properties.[1-5] SGS materials are being investigated for spintronic devices due to their unique magnetic and electrical properties. Recent theoretical work on SGS Heusler compounds in the inverse-Heusler structure has predicted that several of these materials can be antiferromagnetic (AF) gapless semiconductors and half-metallic antiferromagnets (HMAF; also known as fully-compensated half-metallic ferrimagnets).[6,7] Gapless semiconductors are a class of materials where the conduction and valence bands are separated by <0.1 eV band gap. Examples of gapless semiconductors include graphene, HgTe, α-Sn and some half-metallic Heusler compounds.[8-10] Figure 1(a) illustrates the electronic density of states (DOS) for the majority and minority spin components in a gapless semiconductor, where the gap for both the conduction and valence states vanishes at the Fermi energy. In the case where one of the spin bands develops a gap, shown in Fig. 1(b), the electronic system becomes spin-polarized and is referred to as a SGS.

Although the superconducting properties of V$_3$Al in the A15 structure have been known for several decades,[10-13] when synthesized in the pseudo-Heusler D0$_3$ structure V$_3$Al is expected to be an AF gapless semiconductor.[2,6,7] The D0$_3$ structure has the same space group as the L2$_1$ structure, but the binary D0$_3$ structure has a X$_3$Z basis, while the ternary L2$_1$ structure has a X$_2$YZ basis. Figure 1(c) illustrates the D0$_3$ structure of V$_3$Al where the Al atom occupies the 4a Wyckoff position at (0,0,0), the V atom labeled V1 occupies the 4b position at (1/2, 1/2, 1/2), and the remaining two V atoms labeled V2 occupy the 8c position at (1/4, 1/4, 1/4).[11] The atoms along the cubic body diagonal have an Al-V2-V1-V2-Al configuration. The local environment of V1 includes eight V2 nearest-neighbors and six Al second-nearest-neighbors, whereas the V2 atoms have four Al and four V1 nearest-neighbors and six V2 second-nearest-neighbors. It is important to note here that the two V2 atoms are equivalent in the paramagnetic (PM) or ferromagnetic (FM) phases, but it is clear that this symmetry is not preserved in the AF form of this phase. The metastable A15 phase (not shown) is superconducting below ~8-16 K and has a Curie temperature ~20-35 K.[12,13] This A15 structure is unstable between 600-700 $^{o}$C,[14] and previous studies have reported a body-centered-cubic *(bcc)*-type structure forming in this temperature range,[12] but a D0$_3$ structure was not identified.

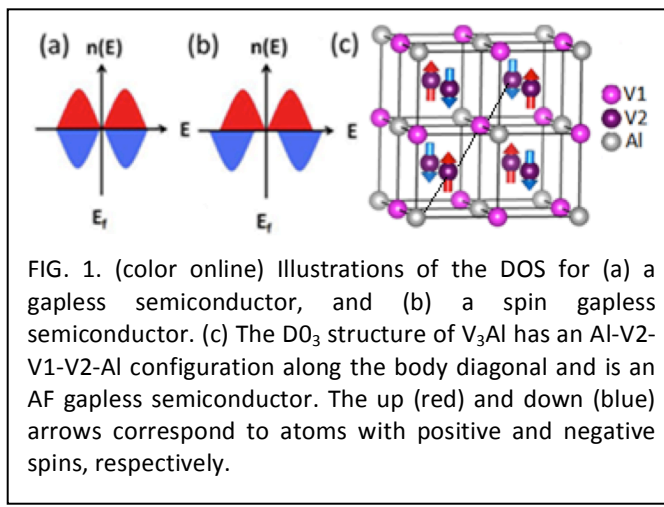

FIG. 1. (color online) Illustrations of the DOS for (a) a gapless semiconductor, and (b) a spin gapless semiconductor. (c) The D0$_3$ structure of V$_3$Al has an Al-V2-V1-V2-Al configuration along the body diagonal and is an AF gapless semiconductor. The up (red) and down (blue) arrows correspond to atoms with positive and negative spins, respectively.

## II. DENSITY FUNCTIONAL THEORY COMPUTATIONS

Total energy and electronic structure calculations were performed using the Quantum Espresso code [15], which is an implementation of density functional theory (DFT) based on the pseudopotential plane-wave method. The calculations were performed using ultrasoft pseudopotentials [16] with plane-wave and charge-density cutoffs of 60 Ry and 400 Ry, respectively. Monkhorst-Pack [17] grids of 20x20x20 and 40x40x40 for the first Brillouin zone sampling were employed for total energy and DOS calculations, respectively. Exchange and electronic correlation have been accounted for using the generalized gradient approximation (GGA-PBE)[18].

Figure 2(a) shows the total energy per formula unit (f.u.) versus unit cell volume for the PM-A15 and PM-D0$_3$, FM-A15, and AF-D0$_3$ phases.[19] The most stable phase appears to be the FM-A15 structure with a modest spin-polarization of 1.4 $\mu_B$/f.u. The energy stabilization due to the spin polarization in this phase is about 9 meV/f.u. The total energy of the PM-D0$_3$ phase lies at a much higher energy, 155 meV above that of the FM-A15 phase. However, within this crystallographic structure the AF order (AF-D0$_3$) reduces the total energy to only 18 meV above the FM-A15 phase. The optimized lattice constant for the AF-D0$_3$ phase is *a* = 6.092 Å.

The AF ordering found in the D0$_3$ structure corresponds to antialigned moments of 1.75 $\mu_B$/atom for the V2 atoms as shown in Fig. 1(c), whereas the V1 atom remains nonmagnetic. As it could be anticipated from symmetry arguments, this AF ordering of the V2 moments in the V2-V1-V2 trimer prevents spin-polarization of the central V1 atom. It was not possible to calculate a FM ground state in the D0$_3$ lattice, as the only non-trivial magnetic order obtained after achievement of self-consistency corresponds to this specific AF order.

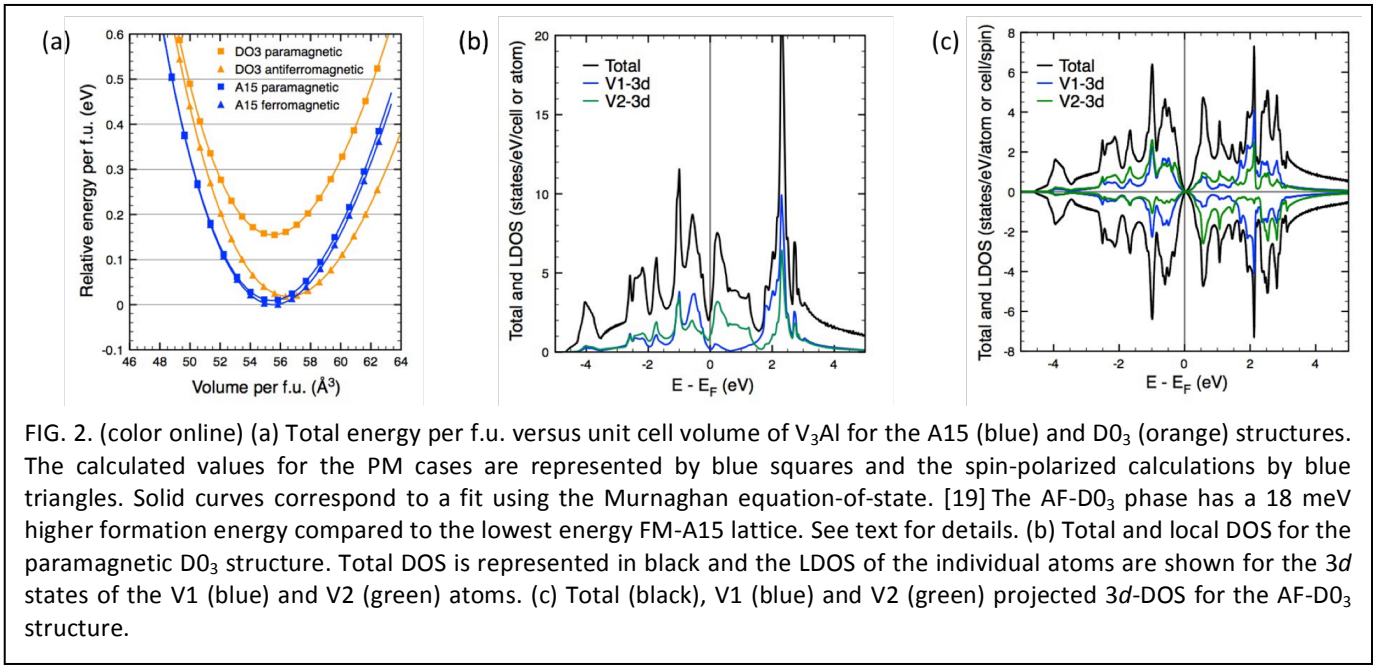

FIG. 2. (color online) (a) Total energy per f.u. versus unit cell volume of V$_3$Al for the A15 (blue) and D0$_3$ (orange) structures. The calculated values for the PM cases are represented by blue squares and the spin-polarized calculations by blue triangles. Solid curves correspond to a fit using the Murnaghan equation-of-state. [19] The AF-D0$_3$ phase has a 18 meV higher formation energy compared to the lowest energy FM-A15 lattice. See text for details. (b) Total and local DOS for the paramagnetic D0$_3$ structure. Total DOS is represented in black and the LDOS of the individual atoms are shown for the 3*d* states of the V1 (blue) and V2 (green) atoms. (c) Total (black), V1 (blue) and V2 (green) projected 3*d*-DOS for the AF-D0$_3$ structure.

In order to understand the origins of the magnetic behavior, the total and local DOS (LDOS) for all the atoms were calculated for the PM-D0$_3$ and AF-D0$_3$ phase and are shown in Figs. 2(b) and (c), respectively. Supplemental information (SI) provides information concerning the symmetry-resolved PM and AF local densities of the 3$d$ states for the V1 and V2 atoms[20]. A ferromagnetic arrangement of the V2 moments would induce a polarization on the central V1 atom, prohibited by the large bonding-antibonding like gap above the Fermi level visible on the V1 $d$-LDOS (see SI for details). The AF arrangement of the V2 moments leads to a much more stable configuration where the central V1 atom is non-polarized. This magnetic structure can be seen as a type G antiferromagnetism involving only the V2 atoms and is characterized by the symmetrical total DOS shown in Fig. 2(c).

### III. EXPERIMENTAL DETAILS

Samples of V$_3$Al were obtained from polycrystalline ingots (~2 grams) prepared by arc melting in an Ar environment using high-purity (99.999 %) elemental metals. The samples' composition was homogenized by annealing in a furnace at 1000 °C with Ar flow for 24 hours.[21] The sample was then cooled to 650 °C and annealed for 4 days, then quenched to retain the D0$_3$ phase.[14] Scanning electron microscopy (SEM) energy dispersive spectroscopy (EDS) confirmed the composition to be stoichiometric V$_3$Al within a <1 % variation across the ingot. Magnetic characterization was performed using a Quantum Design XL-5 superconducting quantum interface device (SQUID) magnetometer with maximum applied field of 5 T in the temperature range 5-400 K. For higher temperature measurements (300-800 K), a Quantum Design VersaLab vibrating sample magnetometer (VSM) with an applied 2 T field was used. X-ray diffraction (XRD) measurements were performed at the Brookhaven National Laboratory National Synchrotron Light Source (NSLS) X14A beamline using λ=0.7783 Å radiation. X-ray magnetic linear dichroism (XMLD) and X-ray magnetic circular dichroism (XMCD) experiments were performed at the NSLS U4B beamline in total electron yield mode. XMCD probes the total vector-sum moment of all the V atoms, $\Sigma m_i$, corresponding to FM and/or ferrimagnetic moments. On the other hand, XMLD measures the total magnitude of the square of the moments, $\Sigma(m_i^2)$, which enables the measurement of AF atoms.[22,23] The XMLD measurements were performed at zero and 1.5 T fields employing incident radiation angles at ± (10°, 20°, 30°, 40°) with respect to the sample's normal direction at room temperature. Due to the sample's polycrystallinity, it is assumed that the XMLD effect is averaged over all the crystallites. Van der Pauw resistivity measurements were carried out using a conductivity probe modified for use in the SQUID magnetometer.[24]

### IV. RESULTS AND DISCUSSION

#### A. X-ray diffraction

Figure 3 shows the XRD Bragg peaks of V$_3$Al expected for a highly-chemically-ordered D0$_3$ phase. The experimental lattice constant was found to be $a$ = 6.144 ± 0.012 Å, in good agreement with the DFT-calculated lattice constant $a$ = 6.092 Å. The chemical order parameter $S$ was determined from the Bragg peak intensities using

$$S = [I_{(220)}/I_{(440)}]^{1/2}/[I'_{(220)}/I'_{(440)}]^{1/2},$$

where I and I' are the measured and theoretical intensities of the Bragg peaks, 2.74 and 3.17 respectively. The theoretical Bragg peak intensities were determined using the appropriate atomic scattering factors.[25,26] The chemical order parameter calculated for the sample is $S = 0.93 \pm 0.08$, indicating a small amount of chemical disorder. The Williamson-Hall[27] plot $\beta\cos\theta$ vs. $\sin\theta$ in Fig. 3(inset) indicates a crystallite size $D = 51$ nm and the presence of minimal crystalline imperfections producing a low inhomogeneous strain (~0.07 %).

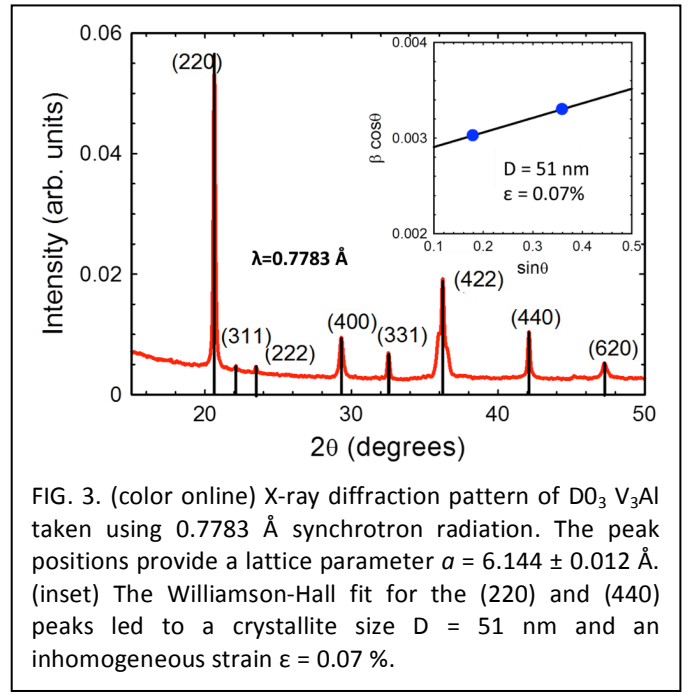

FIG. 3. (color online) X-ray diffraction pattern of D0$_3$ V$_3$Al taken using 0.7783 Å synchrotron radiation. The peak positions provide a lattice parameter $a$ = 6.144 ± 0.012 Å. (inset) The Williamson-Hall fit for the (220) and (440) peaks led to a crystallite size D = 51 nm and an inhomogeneous strain ε = 0.07 %.

**B. Magnetization measurements**

Figure 4 displays a plot of the measured magnetic moment (*m*) in units of $10^{-3}$ $\mu_B$/f.u. as a function of field (*H*) and temperature (*T*). At a field of 2 T the moment was only $10^{-3}$ $\mu_B$/f.u. The inset plots *m*(H) at 300 K and at 600 K, confirming that the magnet moment is strictly linear up to a field of 5 T. Fine structure in the temperature dependence of the moment at 2 T is apparent in Fig. 4, which exhibits a maximum magnetic moment of ~$2.8\times10^{-3}$ $\mu_B$/f.u. at 600 K, consistent with a Néel transition. A similar weak feature in *m*(T) at the Néel transition is typical for polycrystalline antiferromagnets. [28] The existence of antiferromagnetism of $V_3Al$ in the $D0_3$ structure is consistent with: (*i*) the exceedingly small moment at high fields, (*ii*) a nearly temperature-independent moment (varies by <5 %), and (*iii*) the linear variation with field. The small moment arising from AF can be estimated from the model of Van Vleck [29]. Assuming a V moment of 1.64 $\mu_B$ and $T_N=600$ K, the expected moment at 2 T is 0.01 $\mu_B$/f.u., which is in line with measured moment of 0.003 $\mu_B$/f.u. On the other hand, the moment that would arise from Pauli paramagnetism is expected to be much smaller, on order of $10^{-7}$ $\mu_B$/f.u., as expected for transition metals. Thus we can rule out Pauli paramagnetism as the origin of the measured magnetic moment.

**C. X-ray magnetic dichroism**

X-ray magnetic dichroism spectroscopy is convenient for establishing AF in $V_3Al$ by studying polarization differences in the X-ray absorption spectrum (XAS) of the vanadium transitions. Figure 4(b) plots the XAS spectra of the vanadium $L_3$ and $L_2$ transitions along with the polarization-dependent XMCD and XMLD signals. For XMCD, the XAS spectrum did not show any significant change (<0.02%) when comparing data obtained using the right- and left-circular polarizations of radiation, shown by the black curve at zero intensity, indicating the absence of vanadium FM. On the other hand, Fig. 4(b) shows a sizeable change (~$10^{-1}$) in the XAS signal measured using a linearly-polarized x-ray beam. The XAS patterns were collected at various θ angles (±10°, 20°, 30°, 40°) with respect to the perpendicular direction of the sample surface. The peak intensities were normalized to have equal integral areas using the angle-dependent XMLD equation $I(\theta) = I_\parallel \cos^2\theta + I_\perp \sin^2\theta$.[23] Rotation of the sample surface away from the normal direction imparts an out-of-plane intensity component that provides beam intensities that are labeled parallel ($I_\parallel$) and perpendicular ($I_\perp$). Multiple XAS scans were collected over various areas of the sample to ensure that the spectral differences were not due to any chemical differences across the sample. Figure 4(b) plots the intensity of the reflected XMLD signal defined by

$$I_{XMLD} = \frac{I_\parallel - I_\perp}{I_\parallel + I_\perp}.$$

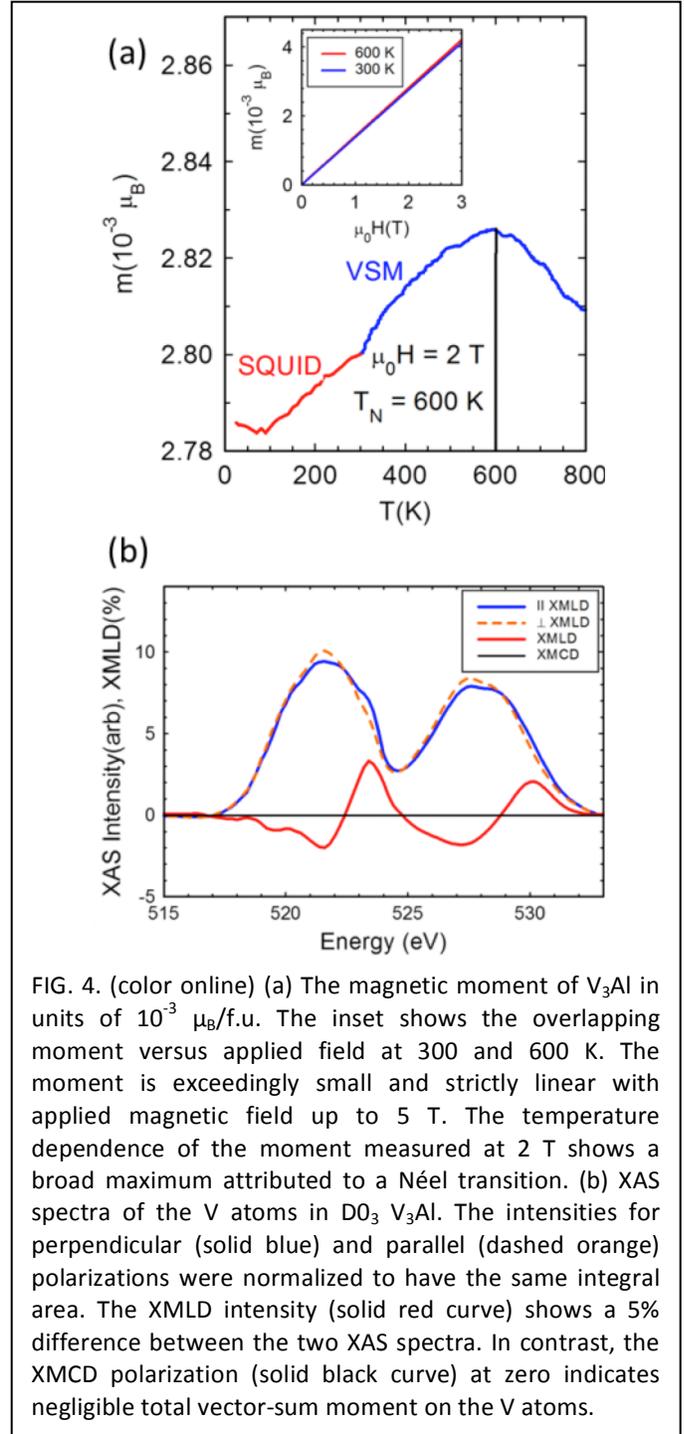

FIG. 4. (color online) (a) The magnetic moment of $V_3Al$ in units of $10^{-3}$ $\mu_B$/f.u. The inset shows the overlapping moment versus applied field at 300 and 600 K. The moment is exceedingly small and strictly linear with applied magnetic field up to 5 T. The temperature dependence of the moment measured at 2 T shows a broad maximum attributed to a Néel transition. (b) XAS spectra of the V atoms in $D0_3$ $V_3Al$. The intensities for perpendicular (solid blue) and parallel (dashed orange) polarizations were normalized to have the same integral area. The XMLD intensity (solid red curve) shows a 5% difference between the two XAS spectra. In contrast, the XMCD polarization (solid black curve) at zero indicates negligible total vector-sum moment on the V atoms.

It can be seen in Fig. 4(b) that the XMLD intensity is ~5 %. Finally, it is noted that 3*d* transition metals generally have a lower spin-orbit interaction energy (~ 50 meV) and a large band dispersion, which reduces the XMLD intensity, but the intensity may be enhanced through multiplet splitting.[23] Generally, transition-metal compounds and alloys show only a moderate XMLD intensity usually in the range 5 - 30 %,[23] and the present measurements fall within this range.

### D. Electrical transport

Electrical measurements were carried out on non-textured $V_3Al$ $D0_3$-phase platelets. The resistivity of the compound was found to be $\rho_{xx}$ = 170 µΩcm at room temperature. The resistivity showed only a small change with temperature (< 4%), consistent with weak phonon contributions to the mobility and constant large carrier concentration. Details of the magnetotransport measurements are provided in the Supplemental section.[20]

## V.    SUMMARY

In summary, DFT calculations confirmed the AF ordering of the compound with only two spin-polarized V atoms out of the three V atoms. By studying the electronic structure of the paramagnetic $D0_3$ phase, we found that the AF order arises from large differences in the local densities of the states close to the Fermi level for the nonequivalent V atoms. Antiferromagnetic $V_3Al$ was synthesized in the $D0_3$ structure via arc-melting and annealing, with its structure and composition confirmed through XRD and EDS measurements. A large XMLD signal was measured in this compound that corresponds to a significant magnetic moment on individual vanadium atoms, but vanishing of the XMCD signal corresponds to an overall AF compensation. SQUID magnetometry confirmed the AF properties through the presence of extraordinarily low magnetization (~$10^{-3}$ $\mu_B$/f.u.) at large fields, and an *m*(T) feature consistent with a Néel transition ~600 K. Further research on the electronic and magnetic properties of this class of antiferromagnetic gapless semiconductors is anticipated to advance the field of nonmagnetic spintronic devices.

**Acknowledgements**

We thank I. McDonald for his assistance with VSM measurements, and F. Jiménez-Villacorta, T. Devakul and A. Feiguin for helpful discussions. The work was supported by the National Science Foundation grants DMR-0907007 and ECCS-1402738. This work was granted access to the HPC resources of IDRIS under the allocations 2014-100384 made by GENCI (Grand Equipement National de Calcul Intensif). Use of the National Synchrotron Light Source (NSLS), Brookhaven National Laboratory, was supported by the U.S. Department of Energy, Office of Science, Office of Basic Energy Sciences, under Contract No. DE-AC02-98CH10886. We thank J. Bai at beamline X14A at NSLS. M.E.J. was supported by the International Centre for Diffraction Data's Ludo Frevel Scholarship.

# Antiferromagnetic phase of the gapless semiconductor V$_3$Al: Supplementary Information

M.E. Jamer, B.A. Assaf, G.E. Sterbinsky, D. Arena, L.H. Lewis, A.A. Saúl, G. Radtke and D. Heiman

## SI. Symmetry-resolved local density of states

Figure S1 shows the symmetry-resolved LDOS for the V1 and V2 atoms in the D03 paramagnetic (PM) and antiferromagnetic (AF) configurations. In the PM state, the different local environments of these two nonequivalent V atoms are clearly reflected in the 3$d$ LDOS. The PM 3$d$ LDOS for V2 shows accessible states at and above the Fermi energy ($E_F$), while the corresponding LDOS for V1 shows a large gap above $E_F$. This gap arises not only from a large covalent interaction with the V2 nearest-neighbors (visible through the large bonding-antibonding-like splitting of the V1 $t_{2g}$ component), but also from the iono-covalent interaction with the Al second-nearest-neighbors (visible through the large bonding-antibonding-like splitting of the V1 $e_g$ component). In the case of the V2 atoms, the existence of states above the Fermi level allows for a spin polarization driven by Hund's intra-atomic exchange, $i.e.$ for a charge transfer from minority spin into higher-lying majority-spin electron states, while the iono-covalent gap prevents it for the V1 atom. Figures S1(c) and (d) show the symmetry resolved LDOS for the AF D03 structure, where it clearly appears that only V2 ions exhibit spin-polarization.

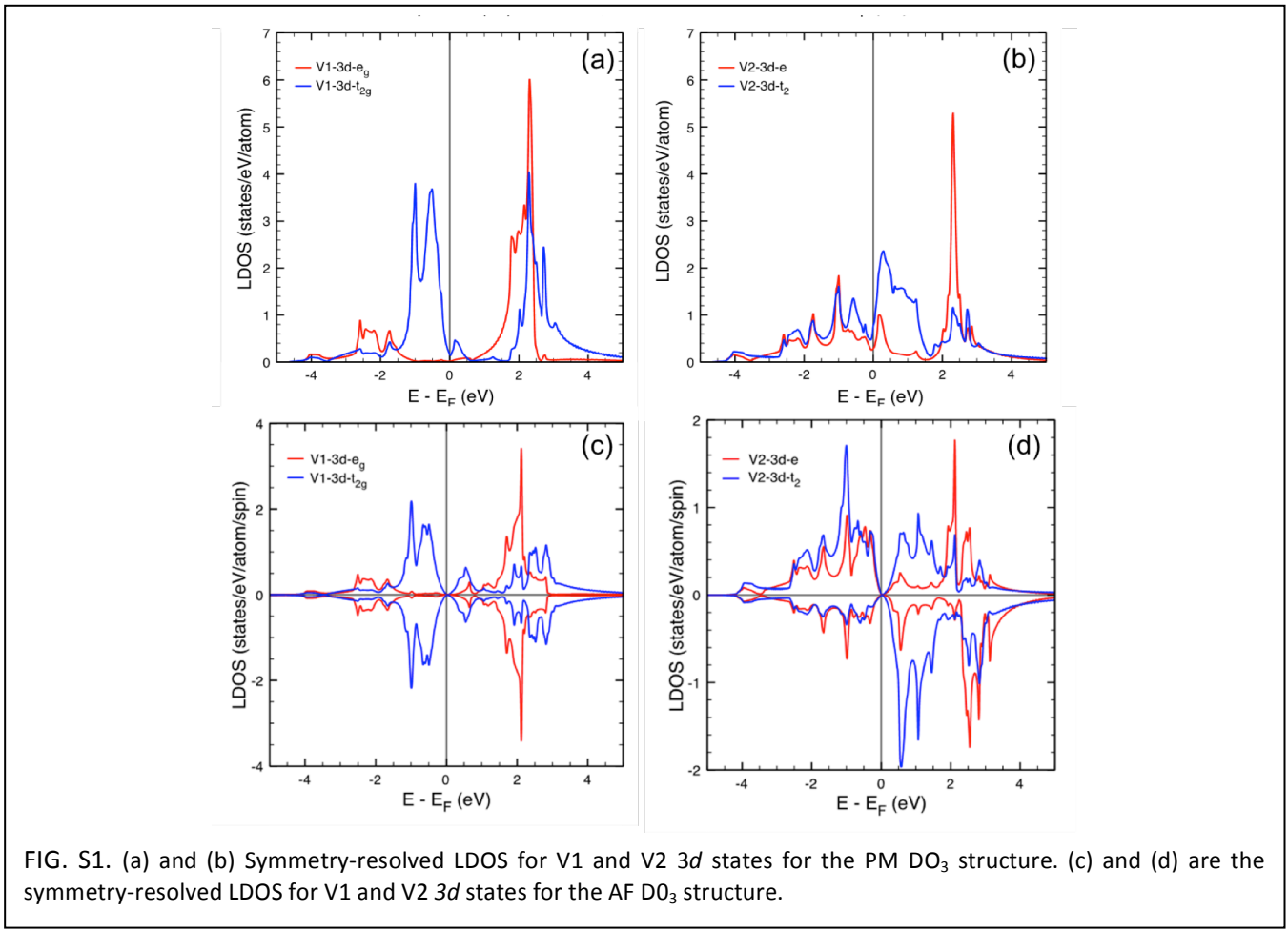

FIG. S1. (a) and (b) Symmetry-resolved LDOS for V1 and V2 3$d$ states for the PM DO$_3$ structure. (c) and (d) are the symmetry-resolved LDOS for V1 and V2 3$d$ states for the AF D0$_3$ structure.

## SII. Resistivity, magnetotransport and magnetization measurements

Figure S2 shows the results of longitudinal resistivity measurements of $V_3Al$ in the range 10-400 K. The temperature dependence $\rho_{xx}(T)$ is quite weak and $\rho_{xx}$ = 170 µΩcm. A remarkable thermal hysteresis was observed in the resistivity as well as the magnetization. It was found that upon heating (red data), $\rho_{xx}(T)$ was independent of the heating rate. In contrast, upon cooling (blue data), $\rho_{xx}(T)$ was a strong function of the cooling rate (the cooling rate 10 K/min is shown), leading to the observed hysteresis. The left inset of Fig. S2 shows the resistance as a function of time for both heating and cooling conditions. The resistivity after heating from 200 to 300 K at 10 K/min (red curve) remained relatively constant (<0.3 % change). On the other hand, the resistivity after cooling from 400 to 300 K at 10 K/min (blue curve) increased by ~15 % over time, having a long exponential relaxation time, $\tau_R$ ~ 10 min. A similar thermal hysteresis in the resistivity behavior has been reported for AF chromium, which was attributed to the presence of AF domain wall motion.[i]

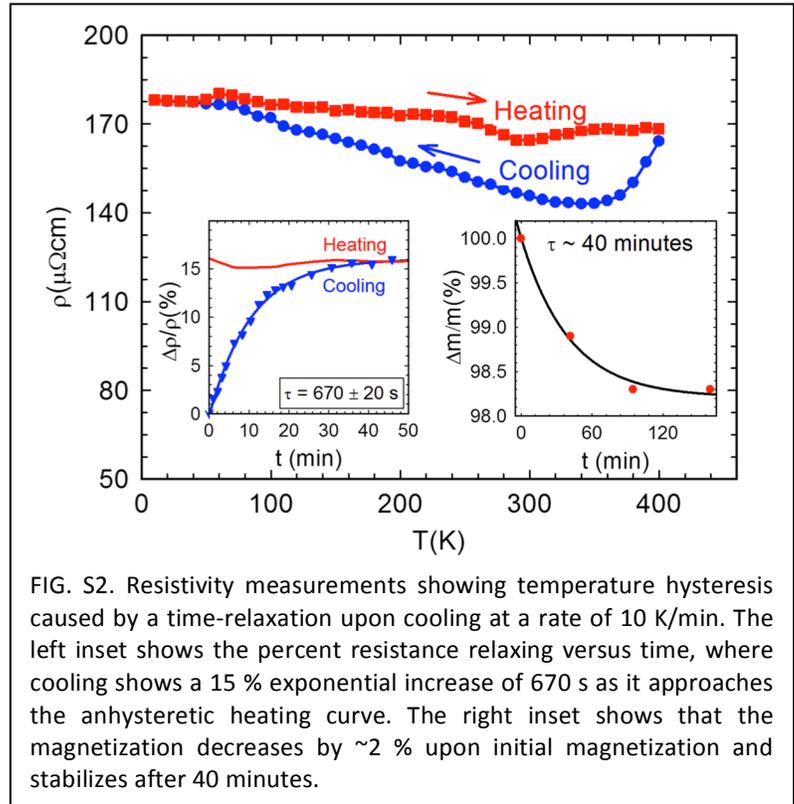

FIG. S2. Resistivity measurements showing temperature hysteresis caused by a time-relaxation upon cooling at a rate of 10 K/min. The left inset shows the percent resistance relaxing versus time, where cooling shows a 15 % exponential increase of 670 s as it approaches the anhysteretic heating curve. The right inset shows that the magnetization decreases by ~2 % upon initial magnetization and stabilizes after 40 minutes.

Furthermore, thermal hysteresis and relaxation was also observed in magnetization experiments. After changing the temperature at a rate of 10 K/min at a magnetic field of 2 T, $m$ was also found exhibit a time relaxation, where $m$ decreased by a few percent. The right inset of Fig. S2 show the slower relaxation having an exponential time constant of $\tau_m$ ~ 40 min. A similar relaxation in the magnetization has been reported in AF $TbFeO_3$ and has been assigned to domain wall motion.[ii] The phenomena observed here in the transport and magnetic data in $D0_3$ $V_3Al$ is consistent with time-dependent AF domain wall motion. Further characterization is beyond the scope of the current investigation.

The magnetoresistance, MR = [R($H$)-R(0)]/R(0), was found to be negligible in applied magnetic fields up to 5 T. This leads to an estimation of the upper limit on the carrier mobility of $\mu \leq 30$ $cm^2$/V, using the classical result, MR = $1 + \mu^2 H^2$, and assuming a small Hall resistivity $\rho_{XY} \ll \rho_{XX}$. [iii] The carrier concentration would then be estimated to be ≥ 1020 carriers/$cm^3$. Such a high carrier concentration is expected to be a consequence of a vanishing band gap coupled with imperfections that shift the Fermi level into the conduction or valence bands. The weak $\rho_{xx}(T)$ dependence and large number of carriers is characteristic of a degenerate semiconductor [iv] and is a feature of the gapless nature of the semiconductor. [v]